\documentclass[
pra,twocolumn,showpacs,preprintnumbers,amsmath,amssymb]{revtex4}

\usepackage{graphicx}
\usepackage{amssymb}
\usepackage{amsmath}
\usepackage{dsfont}
\usepackage{bm}
\usepackage{mathrsfs}
\usepackage{times}
\allowdisplaybreaks

\begin{document}
\title{A note on conserved quantities for electromagnetic waves}

\author{T.\ G.\ Philbin}

\email{t.g.philbin@exeter.ac.uk}

\affiliation{Physics and Astronomy Department, University of Exeter,
Stocker Road, Exeter EX4 4QL, United Kingdom}

\begin{abstract}
Electromagnetic waves carry an infinite number of conserved quantities. We give a simple explanation of this fact, which also shows how to write down conserved quantities at will and calculate their associated symmetry transformations. This framework is then used to discuss decompositions of optical angular momentum, and to prove that magnetic helicity is conserved for beams and pulses. Finally we describe an infinite set of electromagnetic conserved quantities that corresponds to the Virasoro generators of conformal field theories. In the quantum case the Virasoro generators acquire a central charge in their algebra, an example of a quantum anomaly.
\end{abstract}

\maketitle

\section{Introduction}
Every dynamical system with $2n$ canonical variables has $2n$ conserved quantities~\cite{gold}. The proof of this is almost trivial: the initial values of the coordinates and momenta are conserved quantities. In more detail, the solutions $q_i=q_i(q_{0j},p_{0k},t)$ and $p_i=p_i(q_{0j},p_{0k},t)$ for the canonical variables are a set of simultaneous equations for the initial values $\{q_{0i},p_{0i}\}$, and solving these we obtain $2n$ conserved quantities $q_{0i}=q_{0i}(q_{j}(t),p_{k}(t),t)$ and $p_{0i}=p_{0i}(q_{j}(t),p_{k}(t),t)$. Each conserved quantity generates a canonical transformation that leaves the Hamiltonian invariant~\cite{gold}. These canonical transformations are given by the Poisson brackets of the canonical variables $\{q_{i}(t),p_{i}(t)\}$ with the conserved quantities; since the conserved quantities are given as functions of the canonical variables, the transformation can be computed using the canonical Poisson brackets. (After quantization, commutators replace Poisson brackets in this description). Note that this result places no restriction on the Hamiltonian, which will always have the $2n$ symmetries generated by $\{q_{0i},p_{0i}\}$. In general the conserved quantities $q_{0i}=q_{0i}(q_{j}(t),p_{k}(t),t)$ and $p_{0i}=p_{0i}(q_{j}(t),p_{k}(t),t)$ will have an explicit time dependence.

From the above it should be clear that field theories like electromagnetism have an infinite number of conserved quantities. The discrete index on the canonical variables $\{q_{i}(t),p_{i}(t)\}$ of particle mechanics becomes a continuous spatial position label for the canonical variables of a field, giving an uncountably infinite number of canonical variables. This shows that electromagnetic waves have an uncountably infinite number of conserved quantities, given by the initial values of the canonical variables. 

The above argument is intended as a simple explanation of why Maxwell's theory has an infinite number of conserved quantities, a fact that sometimes causes surprise. The rigorous  study of conserved quantities in dynamical systems, including field theories, is a well-developed topic of mathematical physics with strong connections to group representation theory (see for example~\cite{anc01,anc05} for the specific case of Maxwell's equations). This literature is not very approachable for most researchers in optics, yet there is increasing interest in various conserved field quantities in optics and related communities. Examples of current activity in this area are proposals for quantities that measure chirality or helicity of electromagnetic fields~\cite{lip64,mor64,oco64,kib65,can65,cal65,des76,sud86,kri89,prz94,tru96,afa96,ibr09,tan10,hen10,yan11,tan11,bli11,hen12,sch12,ros12,and12,col12,bar12,cam12,cam12b,bli12,bli12b,fer12,fer13,phi13,cam14,cam16,bar18,kel18,lek18,anc01}, and decompositions of field angular momentum into sums of other conserved quantities (see~\cite{lea14,bli14,bli15,bar16} and references therein). The purpose of this Note is to give a simple quantitative presentation of the infinite set of electromagnetic conserved quantities whose existence was deduced above. We then apply this presentation to some widely discussed conserved quantities in electromagnetism. Finally we use our formulation to define a countably infinite set of optical conserved quantities that is well known in the context of (1+1)-dimensional conformal field theories (it is the set of Virasoro generators).

Before beginning the detailed presentation it should be acknowledged that the question of which conserved quantities are useful or interesting is ill defined, and the answers given will to a considerable extent reflect personal taste. There are of course a few conserved quantities (energy, momentum and angular momentum) whose significance is hardly questionable. But some other conserved quantities receive attention only in certain communities. In the general theory of dynamical systems, conserved quantities play a key role in the question of integrability, and such quantities can also generate some nice symmetry groups. For example, six conserved quantities for \emph{nonrelativistic} Coulomb scattering generate the Lorentz group~\cite{gold}. But while this last fact is interesting to many, it is not universally regarded as enlightening. Conserved quantities that generate important symmetries are also not always viewed as significant. For example, Lorentz boost invariance is central in relativistic field theories, but the associated conserved quantity has an explicit time dependence~\cite{wei} and is rarely mentioned. This paper is aimed at the optics and electromagnetism community and assumes its researchers will find it useful to have a simple account of why there are so many optical conserved quantities, and how conserved quantities can be easily written down at will. We also hope that the occurrence in electromagnetism of the Virasoro algebra and its quantum anomaly will be of interest.

%%%%%%%%%%%%%%%%%%%%%%%%%%%%%%%%%%%%%%%
\section{Conserved quantities of electromagnetic waves}
In the Coulomb gauge the general solution for electromagnetic waves is given by the vector potential
\begin{align}
\bm{A}(\bm{r},t)=\int \!\! dV_k & \left(\frac{\xi}{16\pi^{3}\varepsilon_0\omega}\right)^{1/2}\left[ \bm{a}(\bm{k})e^{i(\bm{k\cdot r}-\omega t)} \right. \nonumber  \\
      &  \left.   + \bm{a}^*(\bm{k})e^{-i(\bm{k\cdot r}-\omega t)} \right],  \label{A}  \\
  &   \!\!\!\! \!\!\!\!   \omega=ck, \qquad \bm{k\cdot} \bm{a}(\bm{k})=0,   \label{ak} 
\end{align}
where $dV_k$ is the volume element in $\bm{k}$-space, $\xi$ is a constant with dimensions of action, and each $\bm{a}(\bm{k})$ is the complex amplitude of a vector plane wave with transverse polarization. In the quantum case $\xi$ is replaced by $\hbar$, the complex amplitudes $\bm{a}(\bm{k})$ by annihilation operators, and complex conjugation by Hermitian conjugation. The electric and magnetic fields are $\bm{E}=-\partial_t\bm{A}$ and $\bm{B}=\bm{\nabla\times A}$, respectively.

As discussed in the Introduction, the initial data for electromagnetic waves is a set of conserved quantities. This data is conveniently given by the complex amplitudes $\bm{a}(\bm{k})$, each of which gives the initial amplitude and phase of the plane-wave component with wave vector $\bm{k}$ for two independent polarizations. Each plane wave is essentially a harmonic oscillator and the two real numbers giving the initial values of the oscillator's canonical variables are combined in one complex amplitude.  The $\bm{a}(\bm{k})$ are thus an uncountably infinite set of conserved quantities for electromagnetic waves, and one can write down conserved quantities at will by combining the $\bm{a}(\bm{k})$ together in any of an infinite number of possibilities. The familiar and less familiar optical conserved quantities can be written down in this fashion. 

Moreover each conserved quantity generates a symmetry transformation of the Hamiltonian for the source-free Maxwell theory. The Hamiltonian formulation in Coulomb gauge is well-described in~\cite{wei}, for example. The canonical coordinates and momenta are $\bm{A}$ and $-\varepsilon_0\bm{E}$, respectively. Despite some awkwardness due to the transversality of these fields, the usual machinery of Hamiltonian mechanics can be developed. The canonical Poisson brackets (modified to take account of transversality) generate the Poisson brackets for the amplitudes $\bm{a}(\bm{k})$ and $\bm{a}^*(\bm{k})$.  To obtain the transformation generated by any of the conserved quantities described above, we simply compute its Poisson bracket with any field quantity whose transformation we desire. For example, $\bm{a}(\bm{k})$ itself generates a transformation of $\bm{A}$ that is just an additive displacement by a term proportional to $e^{-i(\bm{k\cdot r}-\omega t)}$. Although this displacement is complex, it is essentially a combination of two real transformations generated by the real and imaginary parts of $\bm{a}(\bm{k})$ (the quadratures for the two plane-wave modes with wave vector $\bm{k}$).

Each of the conserved quantities can be expressed in terms of the gauge-invariant $\bm{E}$ and $\bm{B}$ fields. It is straightforward to invert (\ref{A}) to find the complex amplitudes $\bm{a}(\bm{k})$ in terms of spatial integrals of $\bm{A}$ and $\bm{E}$, and $\bm{A}$ in turn can be written as a spatially nonlocal expression in terms of $\bm{B}$. (For clarity we note that non-locality of a field quantity in $\bm{E}$ or $\bm{B}$ means that its value at point $\bm{r}$ depends on the values of $\bm{E}$ or $\bm{B}$ at other points $\bm{r'}$.)  Since the original conserved quantity is a constant with no $\bm{r}$-dependence, its expression in terms of $\bm{E}$ and $\bm{B}$ will involve at least one (and in general many) spatial volume integrals. Thus the conserved quantity can always be written as an integral $\int dV\,\rho(\bm{r},t)$ over a density $\rho(\bm{r},t)$ that in general is a functional of $\bm{E}$ and $\bm{B}$ involving further volume integrals. Despite $\rho(\bm{r},t)$ having a time dependence, which comes from the time dependence of $\bm{E}$ and $\bm{B}$, its volume integral will be a constant. The scalar field $\rho(\bm{r},t)$ acts a density for the conserved quantity but it must be remembered that in general it is highly nonlocal in $\bm{E}$ and $\bm{B}$, in which case integrals of $\rho(\bm{r},t)$ over \emph{finite} volume regions cannot really be viewed as the amount of the conserved quantity that is present in those regions. Although ``most" optical conserved quantities give a density nonlocal in $\bm{E}$ and $\bm{B}$, there are an infinite number with densities local in $\bm{E}$ and $\bm{B}$~\cite{anc01}. For the latter a flux density $\bm{S}$, also local in $\bm{E}$ and $\bm{B}$, can be identified, giving a local continuity equation $\partial_t\rho+\bm{\nabla\cdot S}=0$~\cite{anc01}.

%%%%%%%%%%%%%%%%%%%%%%%%%%%%%%%%%%%%%%%
\section{Notation}  \label{sec:not}
The representation (\ref{A}) and (\ref{ak}) of an electromagnetic wave shows that in $\bm{k}$-space it corresponds to a vector field $\bm{a}(\bm{k})$ that is everywhere transverse to the radial direction. It is therefore convenient to use spherical polar coordinates in $\bm{k}$-space, which we denote by $\{k,\vartheta,\varphi\}$. Cartesian coordinates in $\bm{k}$-space are aligned with those in $\bm{r}$-space so that $k_z=k\cos\vartheta$, for example. We use a coordinate basis $\{\bm{e}_k,\bm{e}_\vartheta,\bm{e}_\varphi\}$, not a normalized basis, so that $\bm{e}_k\bm{\cdot}\bm{e}_k=1$, $\bm{e}_\vartheta\bm{\cdot}\bm{e}_\vartheta=k^2$ and $\bm{e}_\varphi\bm{\cdot}\bm{e}_\varphi=k^2\sin^2\vartheta$. The metric tensor is $g_{ab}=\text{diag}(1,k^2,k^2\sin^2\vartheta)$ with inverse $g^{ab}=\text{diag}(1,k^{-2},(k\sin\vartheta)^{-2})$. Indices are raised and lowered with $g^{ab}$ and $g_{ab}$, respectively. The vector field $\bm{a}(\bm{k})$ has components $\{0,a^\vartheta(\bm{k}),a^\varphi(\bm{k})\}$ in spherical polars, with nonzero  one-form components $a_\vartheta(\bm{k})=k^2a^\vartheta(\bm{k})$ and $a_\varphi(\bm{k})=k^2\sin^2\vartheta\, a^\varphi(\bm{k})$. The nonvanishing Christoffel symbols are
\begin{gather}
\Gamma^\vartheta_{\ k\vartheta}= \Gamma^\vartheta_{\ \vartheta k}=\Gamma^\varphi_{\ k\varphi}= \Gamma^\varphi_{\ \varphi k} =\frac{1}{k},  \\
\Gamma^k_{\ \vartheta\vartheta}=-k,  \qquad \Gamma^\varphi_{\ \vartheta\varphi}= \Gamma^\varphi_{\ \varphi\vartheta}=\cot\vartheta,  \\
\Gamma^k_{\ \varphi\varphi}=-k\sin^2\vartheta, \qquad  \Gamma^\vartheta_{\ \varphi\varphi}=-\sin\vartheta \cos\vartheta.
\end{gather}
The covariant derivative $\bm{\nabla a}(\bm{k})$ of the vector $\bm{a}(\bm{k})$ is a $\left(\begin{smallmatrix} 1 \\ 1  \end{smallmatrix} \right)$-tensor. Note for example that the component $\nabla_\varphi a^\vartheta(\bm{k})$ of  $\bm{\nabla a}(\bm{k})$ depends on the components $a^\vartheta(\bm{k})$ and $a^\varphi(\bm{k})$ of the vector:  $\nabla_\varphi a^\vartheta(\bm{k})=\partial_\varphi a^\vartheta(\bm{k})-\sin\vartheta \cos\vartheta\, a^\varphi(\bm{k})$.

%%%%%%%%%%%%%%%%%%%%%%%%%%%%%%%%%%%%%%%
\section{Examples}
The energy and momentum of electromagnetic waves are conserved quantities that have well-known expressions in terms of the basic set $\{\bm{a}(\bm{k})\}$:
\begin{gather}
E=\xi \int dV_k\,\omega \bm{a}^*(\bm{k})\bm{\cdot} \bm{a}(\bm{k}), \label{E} \\
\bm{P}=\xi \int dV_k\,\bm{k} \left[\bm{a}^*(\bm{k})\bm{\cdot} \bm{a}(\bm{k})\right].  \label{P} 
\end{gather}
In the quantum case these map to the normal-ordered expressions for the Hamiltonian and momentum operators. These conserved quantities of course generate space-time translations of the electromagnetic fields and they have well-known densities that are local in $\bm{E}$ and $\bm{B}$. In (\ref{E}) and (\ref{P}) all of the basic conserved quantities $\bm{a}(\bm{k})$ appear and the integrals diverge for widely studied waves such as monochromatic beams. In practise this divergence is not a serious issue as one can treat the beam as a very long pulse (which is the situation experimentally). 

The angular momentum vector is less often written explicitly in terms of $\bm{a}(\bm{k})$ (it is given in~\cite{cohen} for example). But usually the interest is in the component of the angular momentum in the direction of propagation of a beam or a pulse, and here a simple expression in terms of $\bm{a}(\bm{k})$ is available but not usually written. We therefore make some brief remarks on a single component of the angular momentum, and also on its decomposition into other conserved quantities

%%%%%%%%%%%%%%%%%%%%%%%%%%%%%%%%%%%%%%%
\section*{Angular momentum}
The angular momentum of a wave in terms of its local density in $\bm{E}$ and $\bm{B}$ is given by $\bm{L}=\varepsilon_0\int dV\bm{r \times}(\bm{E \times B})$~\cite{jac}. If we are interested in only one component of $\bm{L}$ then it is most convenient to choose the $z$-direction and use spherical polar coordinates in $\bm{k}$-space (see section~\ref{sec:not} for notation). The $z$-component of angular momentum  takes a simple form in terms of $\bm{a}(\bm{k})$:
\begin{equation}
L^z=-i\xi \int dV_k \left(  a^*_\vartheta \partial_\varphi a^\vartheta +  a^*_\varphi \partial_\varphi a^\varphi \right),  \label{Lz}
\end{equation}
where we have suppressed the $\bm{k}$-dependence of $\bm{a}(\bm{k})$. This is the $z$-component of angular momentum for any wave but it is most significant when the wave is propagating in the $z$-direction, i.e.\ all the plane-wave components in (\ref{A}) have $k_z>0$. In this case the wave is a beam or a pulse and $L^z$ is the intrinsic angular momentum carried by the wave (see for example~\cite{lea14,bli14,bli15,bar16} for details on optical angular momentum and information on the very large literature on this subject). The option of using derivatives of $\bm{a}(\bm{k})$ in constructing conserved quantities is exploited in (\ref{Lz}).

The expression (\ref{Lz}) shows how the conserved quantity $L^z$ of a wave is determined by its plane-wave components and it allows us to write down waves with any desired value of $L^z$. For example, choosing one of $a^\vartheta$ or $a^\varphi$ to be zero and the other to have a $\varphi$-dependence $e^{im\varphi}$ gives a wave with $L^z=m\xi$ (in the quantum case this gives rise to modes with an $\hat{L}^z$ eigenvalue of $m\hbar$). Note that $m$ does not have to be an integer as there is no requirement for the vector field $\bm{a}(\bm{k})$ in $\bm{k}$-space to be single valued---all that is required is for the integral (\ref{A}) to exist. Of course a dependence $e^{im\varphi}$ with non-integer $m$ can be written as a Fourier series of terms with factors $e^{ip\varphi}$ for all integers $p$.  Waves with a $\varphi$-dependence of $e^{im\varphi}$ in $a^\vartheta$ or $a^\varphi$, with non-integer $m$, are said to have ``fractional orbital angular momentum"~\cite{ber04,lea04,tao05,got08,kot14,tur17,yan18}.

It is important to notice that the derivatives in (\ref{Lz}) are partial derivatives, not covariant derivatives. As noted in section~~\ref{sec:not}, the vector field $\bm{a}(\bm{k})$ in $\bm{k}$-space is always tangent to a sphere centred on the origin and (\ref{Lz}) shows that the angular momentum is determined by how $\bm{a}(\bm{k})$ varies on these spherical surfaces (the dispersion surfaces for the wave). But the angular momentum is independent of the Riemannian connection on the spherical surface because the connection appears in the covariant, not the partial, derivatives. We can form a different conserved quantity by replacing the partial derivatives in (\ref{Lz}) by covariant derivatives, obtaining
\begin{equation}
-i\xi \int dV_k \left(  a^*_\vartheta \nabla_\varphi a^\vartheta +  a^*_\varphi \nabla_\varphi a^\varphi \right).  \label{oam}
\end{equation}
Although not usually written in this form, (\ref{oam}) is the quantity often called ``optical orbital angular momentum" (see~\cite{lea14,bli14,bli15,bar16} for example). It will be noticed that the conserved quantity (\ref{oam}) has the form of the integral of a Berry connection~\cite{nak}. A relationship between some aspects of optical angular momentum and geometric phases is reviewed in~\cite{bli15}. It is interesting that the angular momentum (\ref{Lz}) itself does not involve a nontrivial Berry connection. 

The angular momentum is obtained from (\ref{oam}) by subtracting the connection-dependent terms, i.e.\ by adding the following:
\begin{equation}
-2i\xi \int dV_k \, \cos\vartheta \sin\vartheta \, \mathrm{Im}\left(  a^*_\vartheta a^\varphi    \right).  \label{sam}
\end{equation}
This is also a conserved quantity, often called ``optical spin angular momentum" (see~\cite{lea14,bli14,bli15,bar16} for example). For general waves the densities associated with (\ref{oam}) and (\ref{sam}) are nonlocal in the $\bm{E}$ and $\bm{B}$ fields (in contrast to (\ref{Lz})). The physical significance of the decomposition of the angular momentum into the sum of (\ref{oam}) and (\ref{sam}) is much discussed. We refer the reader to~\cite{lea14,bli14,bli15,bar16} and references therein.

%%%%%%%%%%%%%%%%%%%%%%%%%%%%%%%%%%%%%%%
\section*{Magnetic helicity of beams and pulses}
Consider the conserved quantity
\begin{equation}
-2i\xi \int dV_k \, \sin\vartheta  \, \mathrm{Im}\left(  a^*_\vartheta a^\varphi    \right) , \label{hel}
\end{equation}
which is a simple modification of (\ref{sam}). We now prove that for beams and pulses (\ref{hel}) is the magnetic helicity 
\begin{equation}
h_\mathrm{m}=\varepsilon_0 c\int dV \bm{B\cdot A}.  \label{Bhel}
\end{equation}
Our proof will thus show that magnetic helicity is a conserved quantity for all beams and pulses. 

Magnetic helicity is not conserved for general waves but it \emph{is} conserved if $\int dV \bm{E\cdot B}=0$~\cite{ran89,irv10}. We will derive the condition for conservation of $h_\mathrm{m}$ in  terms of the basic conserved quantities $\bm{a}(\bm{k})$ by computing (\ref{Bhel}) using (\ref{A}). This leads to the expression
\begin{align}
h_\mathrm{m}=-i \int dV_k \,\frac{c\xi}{2\omega} & \left\{ \bm{a}(\bm{k}) \bm{\cdot} \left[ \bm{ k \times a}(-\bm{k}) \right] e^{-2i\omega t}    \right.   \nonumber \\
&  \left.  +  \bm{a}(\bm{k}) \bm{\cdot} \left[ \bm{ k \times a}^*(\bm{k}) \right]  - \mathrm{c.c.} \right\},   \label{hgen}
\end{align}
where $\mathrm{c.c.}$ means complex conjugate. We see that this is $t$-dependent in general so magnetic helicity is not conserved for general waves. Note however that the $t$-dependent term in (\ref{hgen}) contains a product of components of $\bm{a}(\bm{k})$ evaluated at antipodal points on the sphere of radius $k$. A beam or a pulse has all its non-zero complex amplitudes $\bm{a}(\bm{k})$ in a half-space in $\bm{k}$-space because all its plane-wave components have a positive wavenumber with respect to some direction (e.g.\ $k_z>0$). Hence for beams and pulses the time-dependent term in (\ref{hgen}) vanishes. The magnetic helicity is thus conserved in this case and it reduces to (\ref{hel}).

Magnetic helicity is a familiar conserved quantity in plasma physics~\cite{wol58}. It has the remarkable property of giving a measure of the average linkage of field lines of $\bm{B}$~\cite{mof69,chu95,ked16}. This property was discovered when it was shown that in an ideal fluid with velocity field  $\bm{v}$, the conserved quantity $\int dV \bm{v\cdot \nabla\times v}$ measures average linkage of vorticity lines~\cite{mof69}. When magnetic helicity is conserved it thus provides an interesting topological constraint on field dynamics. For electromagnetic waves, conserved magnetic helicity is particularly interesting when it is accompanied by knotted lines of $\bm{B}$. Electromagnetic waves with knotted field lines have been found in the case of pulses~\cite{ran89,irv10,irv08,ked13}, standing waves~\cite{cam18} and monochromatic beams~\cite{phi18}. In the pulse and beam cases the magnetic helicity of knotted $\bm{B}$ fields is conserved. As can be seen from (\ref{Bhel}), the density associated with magnetic helicity is in general nonlocal in $\bm{B}$.

As a consequence of electromagnetic duality, there is an electric helicity with corresponding properties. A combination of electric and magnetic helicity is conserved for all waves~\cite{tru96,afa96,bli12b,bar12,cam12,cam12b}. The associated density is in general nonlocal in $\bm{E}$ and $\bm{B}$.

%%%%%%%%%%%%%%%%%%%%%%%%%%%%%%%%%%%%%%%
\section*{Virasoro algebra in quantum optics}
The previous examples were presented in a classical guise, but essentially the same results hold in the quantum case (modulo operator-ordering issues). We now  give an example of a set of quantum optical conserved quantities that behave differently from their classical counterparts. 

Consider a subset of the basic quantum conserved quantities $\bm{a}(\bm{k})$, namely the subset corresponding to modes with $\bm{k}=n\kappa\bm{\hat{z}}$, where $n$ is a nonzero integer and $\kappa$ is any fixed wave number. In addition we take only one component of $\bm{a}(n\kappa\bm{\hat{z}})$, normalized to give the annihilation operator for these modes. We thus have a discrete set of linearly polarized one-dimensional modes, which would be obtained from periodic boundary conditions in $z$ with period $2\pi/\kappa$. We use the shortened notation $a_n$ for the annihilation operators of this set of modes. A suitable expression for the vector potential (\ref{A}) when the only excitations considered are the restricted set of modes is
\begin{align}
A(z,t)=\sum_{n=-\infty}^{\infty}  &  \sqrt{\frac{\hbar}{4 \pi \varepsilon_0c\mathcal{A} |n|}}
  \left[    a_n e^{i\kappa( n z - c|n| t )}   \right.  \nonumber \\
   & \left.+ a^\dagger_n e^{-i \kappa(n z - c|n| t )} \right],  \label{Adis} 
\end{align}
where $\mathcal{A}$ is a cross-sectional area for the modes. The summation notation in (\ref{Adis}) is a little sloppy as we excluded above the zero mode $n=0$ in the sum; the quantization of this mode would require a term linear in time in $A(z,t)$~\cite{conf}. We have the operator algebra
\begin{equation}
 [a_n,a^\dagger_m]=\delta_{nm},  \qquad  [a_n,a_m]=0,
\end{equation}
where here and in the following it is understood that there are no creation and annihilation operators for $n,m= 0$.

This system of quantum waves is familiar as a $(1+1)$-dimensional conformal field theory~\cite{conf}. It has a set of conserved quantities known as the Virasoro generators, which are of great importance in conformal field theory but are perhaps not so widely recognised in quantum optics. To describe the Virasoro generators we introduce the standard notation~\cite{conf}
\begin{align}
\alpha_n=-i\sqrt{|n|} \left[ \theta(n) a_n - \theta(-n) a^\dagger_{-n}   \right], \\
\widetilde{\alpha}_n=-i \sqrt{|n|} \left[ \theta(n) a_{-n} - \theta(-n) a^\dagger_{n}   \right],
\end{align}
where $\theta(n)$ is the step function (recall that the $n=0$ mode is absent). The main point of these definitions is that $\alpha_n$ ($\widetilde{\alpha}_n$) describe right-going (left-going) waves, as is seen by rewriting (\ref{Adis}):
\begin{align}
A(z,t)=   \sqrt{\frac{\hbar}{4 \pi \varepsilon_0c\mathcal{A}}}    &  \sum_{n=-\infty}^{\infty} \frac{i}{n}
  \left[   \alpha_n e^{-in \kappa(c t - z)}   \right.  \nonumber \\
   & \left.+ \widetilde{\alpha}_n e^{-in \kappa( c t + z )} \right].  \label{Aal} 
\end{align}
We now have the algebra
\begin{gather}
 [\alpha_n,\alpha_m]=n\delta_{n,-m}, \quad [\widetilde{\alpha}_n,\widetilde{\alpha}_m]=n\delta_{n,-m},  \label{alal1}  \\
  [\alpha_n,\widetilde{\alpha}_m]=0,
\end{gather}
without the $m,n=0$ case. For right-going modes, $\alpha_n$ are proportional to annihilation operators when $n>0$ and to creation operators when $n<0$,  with  $(\alpha_n)^\dagger=\alpha_{-n}$. Corresponding results hold for left-going modes and $\widetilde{\alpha}_n$.

The normal-ordered Poynting vector $-\mu^{-1}_0:(\partial_tA)(\partial_zA):$ is an operator that is quadratic in the vector potential (\ref{Aal}), and so it mixes right- and left-going modes. If we include only the right-going modes in the Poynting vector we obtain
\begin{equation}
\frac{\hbar c^2\kappa^2}{2 \pi \mathcal{A}}  \sum_{m=-\infty}^{\infty}L_m e^{-im \kappa(c t - z)},   \label{S} 
\end{equation}
where we have defined
\begin{equation}
L_m=  \frac{1}{2} \sum_{n=-\infty}^{\infty} : \alpha_{m-n} \alpha_n : . \label{vir}
\end{equation}
We have in (\ref{S}) the energy flux operator for right-going modes, written in terms of the \emph{Virasoro generators} $L_m$~\cite{conf}. Although there is no $n=0$ term in the vector potential (\ref{Aal}), the energy flux (\ref{S}) does have a static term from $m=0$ in the sum because there is an $L_0$ operator; this static term in (\ref{S}) is of course the time-averaged energy flux. Note that the normal ordering in (\ref{vir}) is only relevant for $L_0$. It is easy to check that $(L_m)^\dagger=L_{-m}$ so the right-going modes have a Hermitian energy flux (\ref{S}). One defines another set of Virasoro generators $\widetilde{L}_m$ from the $\widetilde{\alpha}_n$, which are relevant for the left-going modes. For readers unfamiliar with $L_m$ it may be useful to see their relation to $a_n$ and $a^\dagger_n$ in a few cases:
\begin{align}
L_0=& \sum_{n=1}^\infty n a^\dagger_n a_n,  \label{L0} \\
L_1=& \sqrt{1.2} \, a^\dagger_1 a_2 +  \sqrt{2.3} \, a^\dagger_2 a_3+ \sqrt{3.4} \, a^\dagger_3 a_4+\dots,  \\
L_2=& -\frac{1}{2} \, a_1 a_1 +  \sqrt{1.3} \, a^\dagger_1 a_3+ \sqrt{2.4} \, a^\dagger_2 a_4+\dots,  \\
L_3=& -\sqrt{2} \, a_1 a_2 +  \sqrt{1.4} \, a^\dagger_1 a_4+ \sqrt{2.5} \, a^\dagger_2 a_5+\dots,
\end{align}

The classical version of (\ref{S}) reveals the basic physical meaning of the Virasoro generators $L_m$: up to a prefactor \emph{they are the complex amplitudes of the frequency components of the Poynting vector for right-going waves}. The Virasoro generators are conserved quantities constructed from the basic conserved quantities $a_n$ of the modes, and they form a closed algebra. To compute the commutators of the Virasoro generators one can appeal to (\ref{alal1}), but there is a subtlety and the correct answer is~\cite{conf}
\begin{equation}
\left[ L_m,L_n  \right] =(m-n) L_{m+n} +\frac{1}{12}(m^3-m)\delta_{m,-n},  \label{val}
\end{equation}
where the final term is a central charge in the algebra. The central charge is a key quantity in conformal field theories and (\ref{val}) is an important example where a quantum relation is \emph{not} obtained from the classical one by mapping Poisson brackets to commutators. The central charge in (\ref{val}) is absent in the classical case and the $L_m$ are then the generatators of the two-dimensional conformal group (in two dimensions the conformal group is infinite dimensional)~\cite{conf}. The central charge is thus an example of a quantum anomaly, where a symmetry of the classical theory (in this case two-dimensional conformal symmetry) is broken in the quantum theory. For this reason the central charge in (\ref{val}) is also called the conformal anomaly~\cite{conf}.

The proof of (\ref{val}) is most elegantly given using complex analysis in an imaginary-time framework~\cite{conf,pol}. The only tricky issue, however, is the central charge in $\left[ L_m,L_{-m}  \right]$, and the correct value for this can be obtained more easily by considering the effect of $L_m$ on the vacuum state~\cite{gsw}. The Jacobi identity applied to commutators of the $L_m$ shows that any central charge in $\left[ L_m,L_{-m}  \right]$ must take the form $c_1m^3+c_2m$, for constants $c_1$ and $c_2$~\cite{gsw}. Moreover it is straightforward to verify that
\begin{equation}
 L_{m}|0\rangle = 0, \quad m\geq -1,  \label{Lm0}
\end{equation}
where $|0\rangle$ is the ground state ($a_n|0\rangle=0$), and also to verify that $\langle 0| L_{2}L_{-2}|0\rangle=1/2$. As a result we find $\langle 0|\left[ L_{1},L_{-1}  \right]|0\rangle=0$ and $\langle 0|\left[ L_{2},L_{-2}  \right]|0\rangle=1/2$, and these two commutator expectation values are enough to determine the two unknowns $c_1$ and $c_2$ in the central charge~\cite{gsw}.

We recall a few more well-known properties of the Virasoro generators~\cite{conf}. They are raising and lowering operators for eigenstates of $L_0$. We see this from (\ref{val}), which gives $\left[ L_0,L_n  \right]=-n L_n$, so if $L_0|h\rangle=h |h\rangle$ then $L_0(L_n|h\rangle)=(h-n)(L_n |h\rangle)$. Moreover this last relation and (\ref{Lm0}) show that $L_{-n}|0\rangle$, $n\geq2$, are (unnormalized) eigenstates of $L_0$, and the central charge appears in the norm of these states: $||L_{-n}|0\rangle||^2=\frac{1}{12}(m^3-m)$. This last equation shows again the necessity of the central charge in the quantum theory: without it the states $L_{-n}|0\rangle$, which are just superpositions of number states, would have zero norm. It can be seen from (\ref{L0}) that there is a high degree of degeneracy in the eigenstates of $L_0$, which are easy to write down as superpositions of number states.

The Poynting vector (\ref{S}) is a wave and by measuring its complex amplitudes we measure the Virasoro generators. As with the electric field, it is convenient to introduce (Hermitian) quadratures as an alternative to the complex amplitudes. The quadratures of the frequency components of the Poynting vector (\ref{S}) are
\begin{equation}
X_m=\frac{1}{2}\left(  L_m+L_{-m}   \right),   \quad Y_m=-\frac{i}{2}\left(  L_m-L_{-m}   \right),     \label{quad}
\end{equation}
where we take only $m\geq 1$ without loss of generality. From (\ref{val}), the quadratures for a given frequency component obey the algebra
\begin{equation}
\left[ X_m,Y_m  \right] = imL_{0} +\frac{i}{24}(m^3-m),  \label{quadal}
\end{equation}
giving the quadrature uncertainty relation
\begin{equation}
\Delta X_m \Delta Y_m \geq \frac{m}{2} \langle L_{0} \rangle + \frac{1}{48}(m^3 - m).   \label{DXY}
\end{equation}
This is a quantum uncertainty relation, but the last term comes from the central charge. Thus the Poynting vector for right-going waves can be said to have a quantum anomaly in its quadrature uncertainty relations.

The Poynting vector's quadratures can in principle be measured and the quantum anomaly in (\ref{DXY}) demonstrated. But the result (\ref{DXY}) is based on a discrete set of modes, and all frequency components of the electric field that contribute to the relevant correlation functions must be included. For example, consider the Poynting-vector quadratures for the vacuum state. The vacuum expectation values of $X_m$ and $Y_m$ are zero but they have vacuum uncertainties $\Delta X_m= \Delta Y_m=\sqrt{(m^3 - m)}/4\sqrt{3}$, and this case satisfies (\ref{DXY}) with equality. For each $m$, this vacuum noise in the quadratues has contributions from a finite number of electric-field frequencies, as can be seen from evaluating $\langle 0| X_m^2|0\rangle$ explicitly using $a_n$ and $a^\dagger_n$. Thus, frequency components that are not excited in the state can still contribute to the relation (\ref{DXY}). There is a somewhat similar phenomenon in quadrature measurements of the electric field using homodyne detection~\cite{lou}. When the local oscillator is at a frequency for which there is no input beam, the resulting measurement can be related to the quadrature uncertainties for the vacuum state~\cite{lou}. The situation with the Virasoro generators is more complex and in terms of the electric field it requires all the relevant  discrete modes, and only those modes, to be probed.

A restriction to a discrete set of modes was made above. For beams with intrinsic angular momentum there is a natural discrete parameter in the modes, namely the integer labelling the angular momentum in the propagation direction. An example is a Bessel beam with an angular dependence $e^{im\phi}$. Here one can choose one frequency for each $m$ to obtain a set of modes with a factor $e^{i(m\phi-|m|\omega_0 t)}$, for some fixed $\omega_0$. We then have a discrete set of modes analogous to that chosen above, but the nontrivial mode functions (involving Bessel functions in this case) are awkward to deal with. An appropriate circular waveguide can be used to simplify the mode functions.

%%%%%%%%%%%%%%%%%%%%%%%%%%%%%%%%%%%%%%%
\section{Concluding remarks}
The existence of symmetries and conserved quantities in dynamical systems is universal and does not necessarily imply special or significant features. We have given a brief account of why electromagnetic waves carry an infinite number of conserved quantities. The treatment here has deliberately avoided any sophisticated mathematical analysis, as this is available elsewhere~\cite{anc01,anc05}. By viewing conserved quantities in terms of their most primitive constituents we gain insight into some well-known electromagnetic quantities such as intrinsic angular momentum and magnetic helicity. The point of view adopted here also allows us easily to see the presence of the Virasoro generators as conserved quantities in electromagnetism.   

It is still not completely clear how many optical conserved quantities can be helpful in the design of structured light for specific purposes or simply to suggest new experiments. Given the enormous number of electromagnetic conserved quantities, this is perhaps not surprising. A range of theoretical tools can be expected to be useful in this area. The machinery used here has the benefit, and the drawback, of being very simple.

%%%%%%%%%%%%%%%%%%%%%%%%%%%%%%%%%%%%%%%
%\section*{Acknowledgements}

%%%%%%%%%%%%%%%%%%%%%%%%%%%%%%%%%%%%%%%%

\end{document}